\documentclass[useAMS,usenatbib]{mn2e}
\usepackage{graphicx,rotate,url,mathptmx,lscape,times,amssymb}
\usepackage{aas_macros}
\usepackage[T1]{fontenc}
\usepackage{aecompl}

\usepackage{color}
\usepackage{amsmath}
\usepackage{amssymb}
\usepackage{multirow}
\font\manual=manfnt at 7pt \def\dbend{\hbox{\raise0.9ex\hbox{\manual\char127\hspace{0.6em}}}}

\newcounter{INTERNALionstage}

\def\gtsim{\mathrel{\hbox{\rlap{\hbox{\lower4pt\hbox{$\sim$}}}\hbox{$>$}}}}
\def\lesssim{\mathrel{\hbox{\rlap{\hbox{\lower4pt\hbox{$\sim$}}}\hbox{$<$}}}}

\def\ga{{\rm\thinspace gauss}}

\def\la{\mbox{{\rm L}$\alpha$}}

\def\hi{\mbox{{\rm H~{\sc i}}}}

\def\hei{\mbox{{\rm He~{\sc i}}}}
\def\heii{\mbox{{\rm He~{\sc ii}}}}

\DeclareMathAlphabet{\vib}{OML}{cmm}{m}{it}

\title[Two-photon continuum of astrophysical nebulae]{
Testing atomic collision theory with the  two-photon continuum of astrophysical nebulae}

\author[F. Guzm\'an et al.]
       {\parbox[]{6.0in}
        { F. Guzm\'an$^1$, N. R. Badnell$^2$, M. Chatzikos$^1$, P. A. M. van Hoof$^3$, R.J.R. Williams$^{4}$ and G.J. Ferland$^{1}$. \\
        \footnotesize
        $^1$Department of Physics and Astronomy, University of Kentucky, Lexington, KY 40506, USA\\
        $^2$Department of Physics, University of Strathclyde, Glasgow G4 0NG, UK\\
        $^3$Royal Observatory of Belgium, Ringlaan 3, 1180 Brussels, Belgium\\
        $^4$AWE plc, Aldermaston, Reading RG7 4PR}        
}

\date{
In preparation}

\pagerange{\pageref{firstpage}--\pageref{lastpage}}
\pubyear{}

\begin{document}

\maketitle

\label{firstpage}

\begin{abstract}

\noindent Accurate rates for energy-degenerate $l$-changing collisions are needed to
determine cosmological abundances and recombination. There are now several competing theories
for the treatment of this process, and it is not possible to test these experimentally.
We show that the \hi{} two-photon continuum produced by astrophysical nebulae is  strongly
affected by $l$-changing collisions. We perform an analysis of the different underlying atomic
processes and simulate the recombination and two-photon spectrum of a nebula containing H and
He. We provide an extended set of effective recombination coefficients and updated
$l$-changing $2s-2p$ transition rates using several competing theories. In principle, accurate
astronomical observations could determine which theory is correct.

\end{abstract}


\section{Motivation}
\label{sec:mot}

Optical recombination lines (ORL) are produced from recombination of ions followed by cascades
from highly excited recombined levels. Theoretical emissivities of these lines should be known
with high accuracy, and observations of such lines in nebulae provide valuable and accurate
information on their composition, temperature and density. This is summarized in the
textbook of \citet{Osterbrock2006}, while \citet{2016PASP..128k4001S} applies this theory to
spectral simulations. 

Although many processes contribute to the formation of these lines, recently attention has
been brought to $l$-changing collisions. These act to redistribute the populations within an
$n$-shell and so change the subsequent cascade to lower levels. Intra $n$-shell $l$-levels are
energy degenerate for hydrogenic systems, so the orbiting electron changes its angular
momentum with no energy change. This is caused by a long range Stark interaction with the
electric field of the projectile. Due to this peculiarity, slow high-mass projectiles are
favored over electrons.

There are now several competing theories for how $l$-changing collisions occur, as summarized
in our previous work \citep{Guzman.I.2016,Guzman.II.2016}, hereafter P1 and P2, and in the
next section. These have an important effect on recombination line intensities, as shown in
P1 and P2. In addition, $l$-mixing collisions in relatively low $n$-shells play an important
role in producing cosmological recombination radiation \citep[CRR;][]{Chluba2010}. In fact,
Ly$\alpha$ and two-photon emissivities are the two bottlenecks at different redshifts:
Ly$\alpha$ at $z>1300$, and two-photon at lower redshift during the recombination epoch.
\citet{Chluba2006} show that an accurate treatment of $l$-mixing collisions, and the 2$s$ and
2$p$ populations is needed to predict the CMB to a precision of $\sim0.1$\%. These processes
have a similar effect on helium recombination lines, which introduces uncertainties in
measurements of the primordial helium abundance \citep{2009MNRAS.393L..36P}.

It is not possible to experimentally determine which of the $l$-changing theories is correct.
Measurements of $l$-changing rates have been done for high-$n$ and low-$l$  Na atoms colliding
with different ions \citep[see, e.g.,][]{MacAdam1980,MacAdam1981,MacAdam1987,Sun1993,Irby1995},
but these levels are not hydrogenic. We know of no experimental measurements of energy
degenerate $l$-changing rates or cross sections. 

In this paper we outline an astronomical observation which, in principle, would make it
possible to discriminate between competing $l$-changing theories. The \hi{} two-photon
continuum, the subject of this paper, is sensitive to the $l$-changing theory since the
population of 2$s$, the level producing the two-photon continuum, is affected by both cascades
from higher levels, and the rate of collisions between 2$s$ and 2$p$. 
Section \ref{sec:2nu_spec} examines collision rates at various temperatures and densities. In
Section \ref{sec:PN}, emissivities are obtained using the different collisional rates. We
present a method to distinguish observationally between the different theoretical methods,
provided that accurate kinetic temperatures are known. Alternatively, if the modified
\citet{PengellySeaton1964} rate coefficients presented in P2 are accepted, the intensity of the
two photon continuum can be used as a diagnostic of the temperature. A discussion of these
results is given in Section \ref{sec:disc}, and a summary is given in Section \ref{sec:sum}.

\section{Determination of $\MakeLowercase{n}=2$ populations and the two photon spectrum} 
\label{sec:2nu_spec}

The \hi{} two-photon continuum emission originates from the transition $2s-1s$, which is
strictly forbidden for single-photon processes. As the total energy of the two photons is
conserved, the energy of one photon can take any value $0 < E < \Delta E_{1s2s}$. The
two-photon process has a spontaneous transition probability which depends on the frequency of
the photons $A_{2s1s,\nu}^{2\nu,H}(\nu)$ \citep{Spitzer1951,NussbaumerSchmutz1984}.

In addition to two photon spontaneous decays to $1s$, the metastable level $2s$ can also be
collisionally excited to $2p$ by $l$-changing transitions and to $np$
($n>2$) by $n$-changing transitions, although the $l$-changing collisions are by far the
fastest process. Such collisions populate 2$p$ which then decays to $1s$ by a dipole
transition producing Ly$\alpha$. Likewise, when the Ly$\alpha$ line is
extremely optically thick in Case B \citep{Brown1970}, l-changing collisions finally transfer
the $2p$ population to $2s$. As a result, the $2s$ population and the two-photon emissivity
will have a strong density dependence. 

The rate coefficients for the $2s - 2p$ $l$-changing transition are tabulated in Table
\ref{t:2s2p} for the competing available theories. The quantum mechanical (QM) approach of
\citet{Vrinceanu2001}, and the Bethe-Born approach of \citet{PengellySeaton1964} (PS64) need a
cut-off to eliminate the divergence of the probabilities at high impact parameters. These
cut-offs depend either on collective effects, such as the plasma Debye length, or the lifetime
of the initial state (P1)\footnote{This can introduce a dependency of the effective rates on
density. However, for the transitions considered in Table \ref{t:2s2p}, the lifetime
cut-off dominates at the astrophysical low densities considered.}. This, together with the
numerical difficulties of the QM calculations at high $n$-shells ($n>60$), due to the lack of 
precision introduced by instabilities caused by the oscillatory behavior of the
hypergeometrical functions needed to compute the cross sections, led \citet{Vrinceanu2001} to
propose a semiclassical (SC) alternative approach. Both QM and SC methods are summarized again
in \citet{VOS2012}. 

P1 shows that the results of the SC approach disagree with the predictions of both the QM and
the PS64 results, mainly because it lacks a complete description of the collision at high
impact parameter, a point made by \citet{2015MNRAS.446.1864S}. Indeed, in Table \ref{t:2s2p},
the QM and modified PS64 (PS-M) approaches give very similar results, which disagree by a
factor $\sim2$ with the SC results. This factor increases up to an order of magnitude for
lower temperatures and up to a factor of $\sim6$ at higher $nl$. The QM and PS64 results
coincide closely with table 4.10 of \cite{Osterbrock2006}, which is called AGN in
Table~\ref{t:2s2p}, and was obtained using PS64. A further simplification, that we have called
the simplified semiclassical method (SSC) to distinguish it from the original SC approach, is
done by \citet{VOS2012} where the scaled angular moment, $l/n$, is taken to be continuous.
Since $n=2$ is such an extreme case, where $l/n$ is highly discrete, it is not surprising that
this approximation does not work well and results are slightly worse. This approximation makes
sense only at sufficiently high quantum number $n$, where the SC predictions may be
approximated by the SSC method to an analytic formula, which is easy to implement. 

The rate coefficients in Table \ref{t:2s2p} decrease with temperature as the Stark $l$-mixing
produced by the charged projectile is less effective at increasing projectile velocities.
As seen in P1, at very low temperatures, the lifetime cut-off decreases the effectiveness of
the collision for QM and PS-M. The SC and SSC methods do not account for these effects
and the rate coefficients continue increasing linearly at low temperatures.

\begin{table*}\footnotesize
\caption{\label{t:2s2p} 
Comparison of rate coefficients ($\text{cm}^{3}\text{s}^{-1}$) by
the different theoretical $l$-changing collision theories for $n=2$ and
at different temperatures. PS-M: Modified method from \citet{PengellySeaton1964} (see P2); QM:
\protect\cite{Vrinceanu2001}; SC: \protect\cite{Vrinceanu2001}; SSC: Simplified SC method of
\protect\cite{VOS2012}; AGN: Results from table 4.10 of
\protect\cite{Osterbrock2006} obtained using PS64.}
\begin{tabular}{c c c c c c c c }
\cline{3-8}
& & \multicolumn{1}{|c|}{$T_\text{H}=100\text{K}$}&\multicolumn{1}{|c|}{$T_\text{H}=1000\text{K}$}
&\multicolumn{1}{|c|}{$T_\text{H}=5000\text{K}$}&\multicolumn{1}{|c|}{$T_\text{H}=10000\text{K}$}
&\multicolumn{1}{|c|}{$T_\text{H}=15000\text{K}$}&\multicolumn{1}{|c|}{$T_\text{H}=20000\text{K}$}\\
\hline
\multicolumn{1}{|c}{\multirow{5}{*}{$2s\,^2\text{S}\to 2p\,^2\text{P}_{1/2}^0$}} 
& \multicolumn{1}{|c|}{QM} & 1.73e-4 & 3.55e-4 & 2.92e-4 & 2.49e-4 & 2.24e-4 & 2.07e-4\\
\cline{2-2}
\multicolumn{1}{|c}{} & \multicolumn{1}{|c|}{PS-M} & 1.55e-4 & 3.58e-4 & 2.96e-4
& 2.52e-4 & 2.26e-4 & 2.09e-4 \\
\cline{2-2}
\multicolumn{1}{|c}{} & \multicolumn{1}{|c|}{SC} & 1.53e-3 & 4.82e-4 & 2.15e-4 
& 1.53e-4 & 1.25e-4 & 1.08e-4 \\
\cline{2-2}
\multicolumn{1}{|c}{} & \multicolumn{1}{|c|}{SSC} & 4.18e-4 & 1.32e-4 & 5.92e-5 
& 4.18e-5 & 3.42e-5 & 2.96e-5 \\
\cline{2-2}
\multicolumn{1}{|c}{} & \multicolumn{1}{|c|}{AGN(PS64)} & -- & -- & -- & 2.51e-4 & -- & 2.08e-4 \\
\hline
\multicolumn{1}{|c}{\multirow{4}{*}{$2s\,^2\text{S}\to 2p\,^2\text{P}_{3/2}^0$}} 
& \multicolumn{1}{|c|}{QM} & 4.92e-6 & 1.17e-4 & 2.24e-4 & 2.30e-4 & 2.26e-4 & 2.20e-4 \\
\cline{2-2}
\multicolumn{1}{|c}{} & \multicolumn{1}{|c|}{PS-M} & 4.56e-6 & 1.06e-4 & 2.24e-4
& 2.34e-4 & 2.28e-4 & 2.22e-4 \\
\cline{2-2}
\multicolumn{1}{|c}{} & \multicolumn{1}{|c|}{SC} & 2.88e-3 & 9.11e-4 & 4.08e-4 
& 2.88e-4 & 2.35e-4 & 2.04e-4 \\
\cline{2-2}
\multicolumn{1}{|c}{} & \multicolumn{1}{|c|}{SSC} & 8.36e-4 & 2.64e-4 & 1.18e-4 
& 8.36e-5 & 6.84e-4 & 5.92e-5 \\
\cline{2-2}
\multicolumn{1}{|c}{} & \multicolumn{1}{|c|}{AGN(PS64)} & -- & -- & -- & 2.23e-4 & -- & 2.19e-4 \\
\hline
\end{tabular}
\end{table*}


\begin{table*}\footnotesize
\caption{\label{t:effreccoef} 
  Comparison of effective recombination coefficients
  ($10^{-13}\text{cm}^{3}\text{s}^{-1}$) to {\bf $2s$} and  {\bf $2p$} calculated for Case A and
  Case B using QM $l$-changing collisions at different temperatures and
  densities. P64: Results from \protect\citet{Pengelly1964} are quoted in
  \protect\cite{Osterbrock2006}.
}
\begin{tabular}{c c c c c c c c c }
\cline{4-9}
& & & \multicolumn{1}{|c|}{$T=100\text{K}$}&\multicolumn{1}{|c|}{$T=1000\text{K}$}
&\multicolumn{1}{|c|}{$T=5000\text{K}$}&\multicolumn{1}{|c|}{$T=10000\text{K}$}
&\multicolumn{1}{|c|}{$T=15000\text{K}$}&\multicolumn{1}{|c|}{$T=20000\text{K}$}\\
\hline
\multicolumn{1}{|c}{\multirow{8}{*}{Case A}}
& \multicolumn{1}{|c}{\multirow{4}{*}{$2s$}} 
& \multicolumn{1}{|c|}{$n_e=n_\text{H}=10^{2}\text{cm}^{-3}$} & 4.16 & 1.15 & 0.475 & 0.318 & 0.249 & 0.208 \\
\cline{3-3}
& \multicolumn{1}{|c}{} & \multicolumn{1}{|c|}{$n_e=n_\text{H}=10^{4}\text{cm}^{-3}$} & 5.77 & 1.22 & 0.481 & 0.320 & 0.250 & 0.209 \\
\cline{3-3}
& \multicolumn{1}{|c}{} & \multicolumn{1}{|c|}{$n_e=n_\text{H}=10^{6}\text{cm}^{-3}$} & 14.9 & 1.45 & 0.503 & 0.328 & 0.254 & 0.211 \\
\cline{3-9}
& & \multicolumn{1}{|c}{\multirow{1}{*}{P64($n_e=n_\text{H}\to0$)}} & -- & -- & 0.475 & 0.318 & -- & 0.208 \\
\cline{2-9}
& \multicolumn{1}{|c}{\multirow{4}{*}{$2p$}} 
& \multicolumn{1}{|c|}{$n_e=n_\text{H}=10^{2}\text{cm}^{-3}$} & 56.2 & 10.8 & 3.02 & 1.65 & 1.14 & 0.869 \\
\cline{3-3}
& \multicolumn{1}{|c}{} & \multicolumn{1}{|c|}{$n_e=n_\text{H}=10^{4}\text{cm}^{-3}$} & 65.3 & 10.6 & 2.98 & 1.64 & 1.13 & 0.864 \\
\cline{3-3}
& \multicolumn{1}{|c}{} & \multicolumn{1}{|c|}{$n_e=n_\text{H}=10^{6}\text{cm}^{-3}$} & 123 & 10.9 & 2.92 & 1.61 & 1.11 & 0.853 \\
\cline{3-9}
& & \multicolumn{1}{|c}{\multirow{1}{*}{P64($n_e=n_\text{H}\to0$)}} & -- & -- & 3.07 & 1.67 & -- & 0.877 \\
\hline
\multicolumn{1}{|c}{\multirow{8}{*}{Case B}}
& \multicolumn{1}{|c}{\multirow{4}{*}{$2s$}} 
& \multicolumn{1}{|c|}{$n_e=n_\text{H}=10^{2}\text{cm}^{-3}$} & 14.1 & 3.54 & 1.33 & 0.835 & 0.625 & 0.505\\
\cline{3-3}
& \multicolumn{1}{|c}{} & \multicolumn{1}{|c|}{$n_e=n_\text{H}=10^{4}\text{cm}^{-3}$} & 20.4 & 3.76 & 1.34 & 0.838 & 0.626 & 0.505 \\
\cline{3-3}
& \multicolumn{1}{|c}{} & \multicolumn{1}{|c|}{$n_e=n_\text{H}=10^{6}\text{cm}^{-3}$} & 53.3 & 4.50 & 1.39 & 0.850 & 0.630 & 0.507 \\
\cline{3-9}
& & \multicolumn{1}{|c}{\multirow{1}{*}{P64($n_e=n_\text{H}\to0$)}} & -- & -- & 1.33 & 0.837 & -- & 0.507 \\
\cline{2-9}
& \multicolumn{1}{|c}{\multirow{4}{*}{$2p$}} 
& \multicolumn{1}{|c|}{$n_e=n_\text{H}=10^{2}\text{cm}^{-3}$} & 59.0 & 11.2 & 3.17 & 1.74 & 1.20 & 0.920 \\
\cline{3-3}
& \multicolumn{1}{|c}{} & \multicolumn{1}{|c|}{$n_e=n_\text{H}=10^{4}\text{cm}^{-3}$} & 72.3 & 11.3 & 3.15 & 1.74 & 1.20 & 0.918 \\
\cline{3-3}
& \multicolumn{1}{|c}{} & \multicolumn{1}{|c|}{$n_e=n_\text{H}=10^{6}\text{cm}^{-3}$} & 152 & 12.2 & 3.18 & 1.74 & 1.21 & 0.922 \\
\cline{3-9}
& & \multicolumn{1}{|c}{\multirow{1}{*}{P64($n_e=n_\text{H}\to0$)}} & -- & -- & 3.21 & 1.76 & -- & 0.927 \\
\hline
\end{tabular}
\end{table*}

\begin{figure}
\begin{center}
\includegraphics[width=0.4\textwidth]{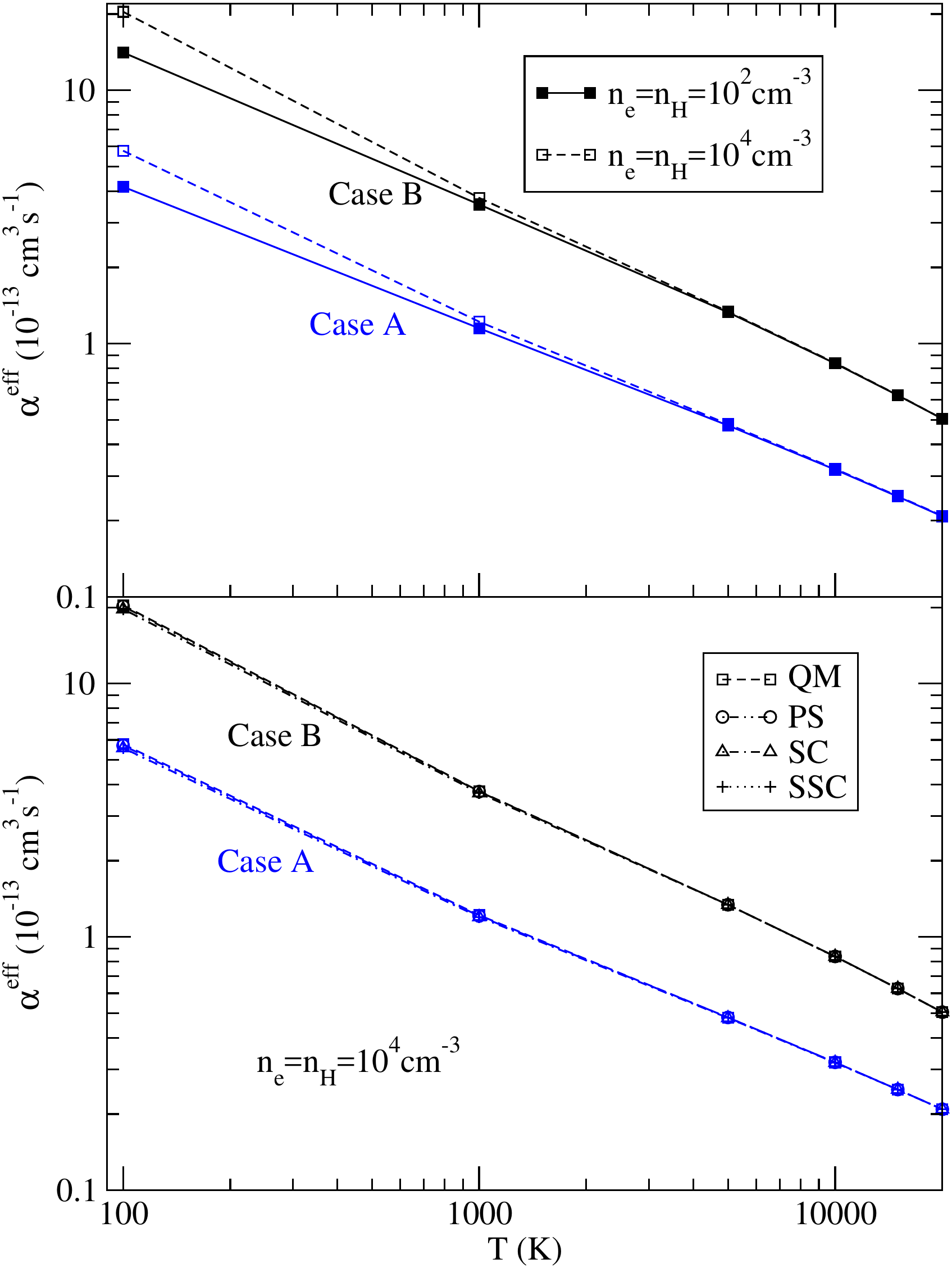}
\caption{\label{f:effreccoef} Top: effective recombination coefficients to $2s$ using QM $l$-changing method for
Case A (blue) and B (black) and different densities. Bottom: comparison of different $l$-changing method for
effective coefficients for Case A (blue) and B (black). }
\end{center}
\end{figure}  

The rates in Table \ref{t:2s2p}, together with the other collisional and radiative processes,
are combined with  effective recombination coefficients to predict the 2$s$ population and
thus the two-photon continuum emissivity. In equilibrium, and in absence of 
charge exchange, the population of the hydrogen $nl$-level, $n_{nl}$, is given by:
\begin{align}
  n_en_p\alpha_{nl}^{eff}+\sum_{n^\prime l^\prime \neq nl}n_en_{n^\prime l^\prime}q_e\left(n^\prime l^\prime \to nl\right)+\sum_{l^\prime \neq l}n_pn_{n l^\prime}q_p\left(nl^\prime \to nl\right) &=\nonumber\\
  n_{nl}\left[\sum_{n^\prime l^\prime \neq nl}n_eq_e\left(nl\to n^\prime l^\prime \right)+\sum_{l^\prime \neq l}n_pq_p\left(nl\to nl^\prime \right)+n_eS_{nl} + \sum_{n^\prime l^\prime < nl}A_{nl \to n^\prime l^\prime}\right] & \, ,
\label{eq:eq}  
\end{align}
\noindent where $A_{nl \to n^\prime l^\prime}$ are the Einstein coefficient for
the spontaneous $nl \to n^\prime l^\prime$ decay; $n_e$, $n_p$
are the electron and proton 
density respectively, $q_e\left(n^\prime l^\prime \to nl\right)$ is the
collisional transition rate (by electron impact) from the $n^\prime l^\prime$-shell
to the $nl$-shell  and $q_p\left(n l^\prime \to nl\right)$ is the
intra $n$-shell collisional transition rate (by proton impact) from the $l^\prime$-subshell to
the $l$-subshell, $S_{nl}$ are the
ionization rate coefficients and $\alpha_{nl}^{eff}$ are the effective
recombination coefficients to the $nl$ shell. Proton n-changing collisions and
electron l-changing collisions are neglected. The effective recombination coefficients are
given by:
\begin{equation}
n_en_p\alpha_{nl}^{eff}=n_en_p\alpha^{rad}_{nl} + n_e^2n_p\alpha^{3b}_{nl} +\sum_{n^\prime l^\prime>nl}n_{n^\prime l^\prime}A_{n^\prime l^\prime\to nl} \, ,
\label{eq:sef}
\end{equation}
where $\alpha^{rad}_{nl}$ are the radiative recombination rate coefficients 
(in units of $\text{cm}^3\text{s}^{-1}$) and  $\alpha^{3b}_{nl}$ three-body recombination rate 
coefficients (in units of $\text{cm}^6\text{s}^{-1}$). $\alpha_{nl}^{eff}$ includes direct 
recombination to the $nl$-shell plus decays from upper levels that have 
already recombined and/or suffered collisions.

We have used the development version of the spectral code 
Cloudy, which is latest most recently described by \citet{CloudyReview}, to calculate  the 
$\alpha_{nl}^{eff}$. The calculated coefficients are given in Table \ref{t:effreccoef} and Fig.
\ref{f:effreccoef} for different temperatures and densities where the QM method has been used.
Case A and Case B \hi{} line formation have been assumed \citep{Osterbrock2006}. Case B
systematically produces larger recombination line emissivities since Ly lines are re-absorbed
and converted into Balmer lines \citep{Baker1938}. At temperatures $T>10^3$~K the effective
coefficients depend weakly on the density and agree with the collisionless work of 
\citet{Pengelly1964}. This is because radiative decays are faster than collisions at these
densities. In Table \ref{t:effreccoef} and the top graph of Fig. \ref{f:effreccoef}, it is
seen that the density dependence  becomes strong at low temperatures $T<10^3$~K. High
$n$-shell $l$-changing collisions are more efficient at these temperatures, producing a more
effective $l$-mixing. This  affects the cascade towards the low levels and changes the
populations. Different $l$-changing methods are also compared in the bottom graph of Fig.
\ref{f:effreccoef} and no differences are found between them. We conclude that
the differences between $l$-changing methods 
have a small influence on effective recombination coefficients even at low temperatures. 
However, once  the metastable  $2s$ level is reached, $l$-changing collisions strongly
influence the $2s$-$2p$ populations, and produce measurable differences in the two-photon
emission.     

\section{Dependence of two photon continuum on temperature and density}
\label{sec:PN}

Using the different theoretical approaches we  run simulations with the spectral code Cloudy
and calculated emissivities for a layer of H/He gas at constant temperature, illuminated by a
mono-energetic ``laser'' radiation of 1.1~Ryd, enough to ionize hydrogen but not to fluoresce
any \hi{} lines. This is done in order to model an optically thick environment which keeps
with case B assumptions \citep{Baker1938}. For simplicity, we assume that the heavy elements
have negligible
abundances, while the He/H abundance ratio is taken to be 10\%,  similar to cosmic abundances.
The ionization parameter, the ratio of ionizing photon to hydrogen atom densities, is U=0.1.
The chosen temperature and density of the hydrogen atoms is $T_e=15000$~K
and $n_\text{H}=10^4 \text{cm}^{-3}$. The electron density is calculated consistently depending
on the He ionization degree (H is fully ionized while He is atomic due to the low energy of
the laser). If He atoms are singly ionized then the electron density would be 10\% higher than
$n_\text{H}$. Our proposed test compares Case B predictions with observations
of a nebula. It is possible that such processes as continuum pumping of the Lyman lines, or their
escape from the cloud, would mitigate the Case B assumption and change the resulting emission.
As a test we recomputed a standard model of a young planetary nebula, including our large H I
model.  The model is our "pn\_paris" in the Cloudy test suite.  It was a standard model in the
1984 Meudon meeting \citep{Pequignot.D86Comparison-of-Photoionization-and-steady-shock}, and
has been included in subsequent photoionization workshops
\citep{Ferland_Lexingtonbenchmark,Pequignot2001}.
We found that the H I lines agreed with the
\citet{Hummer1987} Case B predictions to a small fraction of a percent for H I lines,
$\sim0.3$\%. Our pure Case B calculations agree with the \citet{Hummer1987} ones to about this
difference, due to change in the atomic data over the past few decades. In view of that, we can
rely on Case B approximation as highly accurate. 

\begin{figure}
\begin{center}
\includegraphics[width=0.4\textwidth]{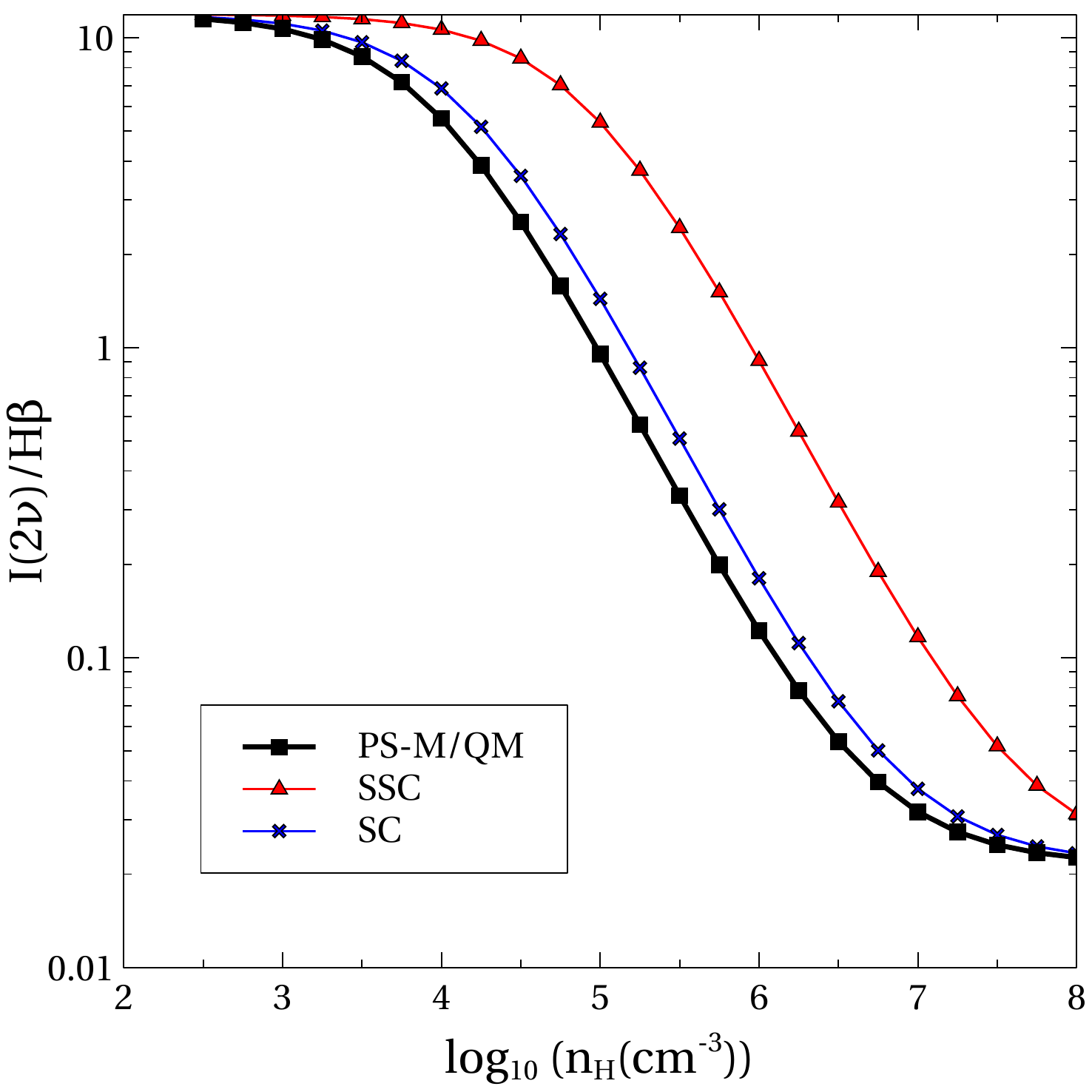}
\caption{\label{f:15K} The total integrated two-photon to H$\beta$ intensity ratio as a
function of hydrogen density. The different lines represent the different theories for the $l$-changing
rates.
}
\end{center}
\end{figure}

The ratio of the total integrated two-photon continuum intensity relative to Case B H$\beta$
is presented in Fig. \ref{f:15K} for the different $l$-changing theories. The relative
two-photon flux decays exponentially for intermediate densities since $l$-changing collisions
depopulate the $2s$ levels by $2s-2p$ transitions before they can radiate. At these densities
the H$\beta$ emissivity ratio $4\pi j$(H$\beta$)$/n_e n_p$ does not vary by much. 
In Fig. \ref{f:15K}, the lower $l$-changing rates from the SSC approach have the effect of
maintaining a higher population in the $2s$ level at intermediate densities, producing a
higher two-photon flux. However, as it has been pointed out in Section \ref{sec:2nu_spec},
this approximation is probably wrong for low $n$-shells with a highly discrete $l/n$ grid. On
the other hand, SC calculations only give a slightly higher emissivity than the PS-M and QM
calculations, which are indistinguishable from each other. The SC approach usually gives rates
that are a factor $\sim6$ lower than PS-M and QM (see P1, P2 and \citet{VOS2012}). However,
for the transition listed in Table \ref{t:2s2p}, the very similar values of the SC and QM rate
coefficients cause the two-photon flux predictions to be similar.
 
Figure \ref{f:15K} shows that the two-photon to H$\beta$ intensity ratio is strongly affected
by both density and collision theory. The two-photon continuum can be detected on the same
spectrum as \hi{} Balmer lines so it should be possible to measure this ratio. It may be
possible to observationally test which theory is correct by observing the continuum at a
wavelength with no contamination from other lines. It is possible to determine the kinetic
temperature and gas density from the emission-line spectrum, as described in
\citet{Osterbrock2006}. The instrumental properties and the density of lines would affect the
choice of the specific wavelength range to conduct this test.

\begin{figure}
\begin{center}
\includegraphics[width=0.4\textwidth]{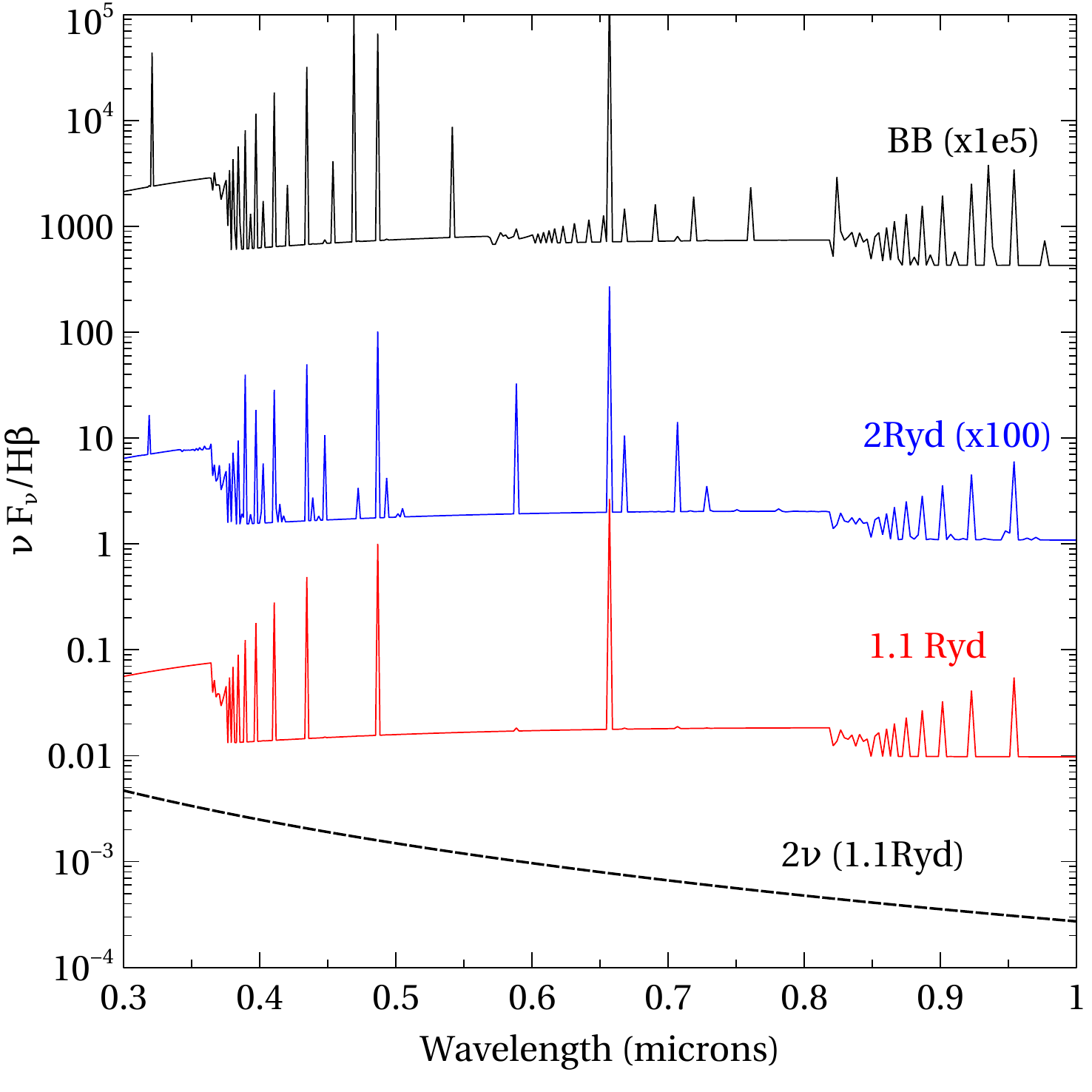}
\caption{\label{f:spectrum} Cloudy simulations of the emission spectrum of a mono-layer of
He and H gas with He/H=0.1 at 15~000K and $n_\text{H}=10^4\text{cm}^{-3}$ illuminated by a
$10^5$~K black body and two mono-energetic ``laser'' radiation fields at 1.1 Ryd and 2 Ryd
respectively (see text). The \hi{} two-photon radiation contribution is shown for the 1.1~Ryd
case (dashed line). The incident continuum has been subtracted from the blackbody spectrum.
The 2 Ryd and the BB spectra have been multiplied by $10^2$ and $10^5$ for clarity.}
\end{center}
\end{figure}

Figure \ref{f:spectrum} shows the full spectrum normalized to the H$\beta$
flux for our pure H and He cloud. We have carried out photoionization simulations with three
different incident radiation fields. The first was a mono-energetic radiation of 1.1 Ryd which
can ionize only H. A second mono-energetic radiation field at 2 Ryd could ionize H and produce
He$^+$. Helium ions are heavier than protons and slower at the same temperature.
$l$-changing rates with He$^+$ ions will be higher than those produced by protons. However,
this contribution is reduced by the much lower abundance of helium atoms. Finally, a black
body (BB) radiation field with temperature $T_{BB}=10^5$~K which can fully
ionize both H and He. This is plotted with the ``BB'' label and corresponds to a
Planetary Nebula. Again, line fluorescence processes have been excluded.
When the incident continuum is bright enough to ionize He, as in the top and middle spectra of
Fig. \ref{f:spectrum}, \hei{} recombination lines from $n>2$ to $2s$ and $2p$ are on the
spectrum. Helium is mostly singly ionized when the 2 Ryd radiation is used and 
double ionized in the case of BB where \heii{} series is also visible. visible.
In these two latter cases the two-photon continuum will be the sum of H and
He contributions. In this work, for simplicity, we will take the case of 1.1 Ryd radiation
which ionizes H only. In our calculations, \hei{} two-photon contribution can be up to a
$\sim15$\% for the 2 Ryd radiation case and is less than 0.2\% for the 1.1 Ryd radiation case.

\begin{figure}
\begin{center}
\includegraphics[width=0.4\textwidth]{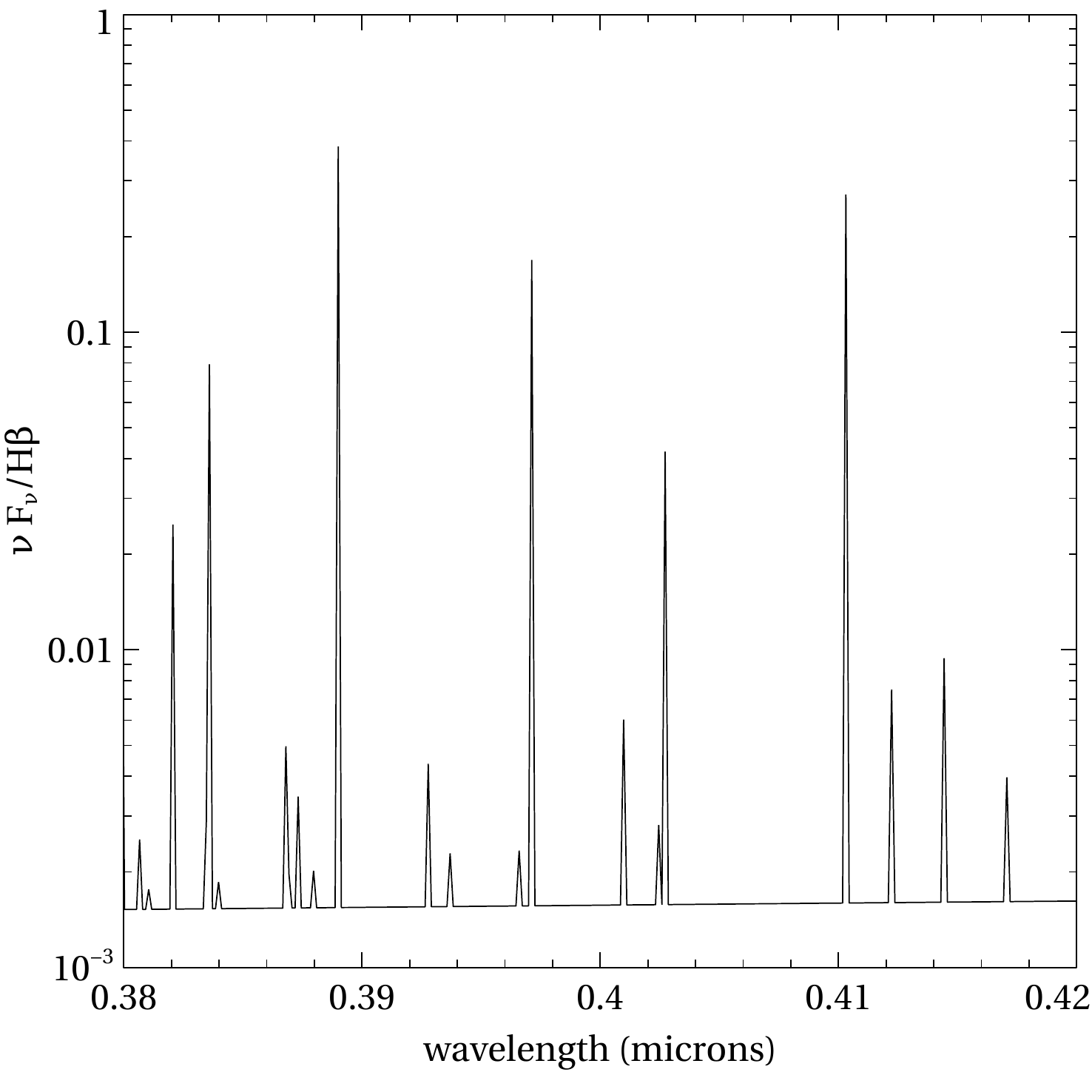}
\caption{\label{f:finespec} Fine resolution Laser 1.1 Ryd spectrum
(see Fig. \ref{f:spectrum}). }
\end{center}
\end{figure}  

It is desirable  to use a wavelength that is dominated by the two photon continuum and clear
of emission lines. Regions just longward of the Balmer Jump (BJ) would be difficult due to the
confluence of the hydrogen Balmer lines. For reference, we have chosen a region around
3900\AA, where no Balmer lines occur (see Fig.  \ref{f:finespec}) and where
the two-photon flux is a major contributor to the total continuum (see
Fig. \ref{f:spectrum}). This
wavelength is in regions where CCD detectors are still efficient. However, due to the smooth
behavior of the two photon continuum any slight alteration of the selected wavelength will
produce the same results.

\begin{figure}
\begin{center}
\includegraphics[width=0.4\textwidth]{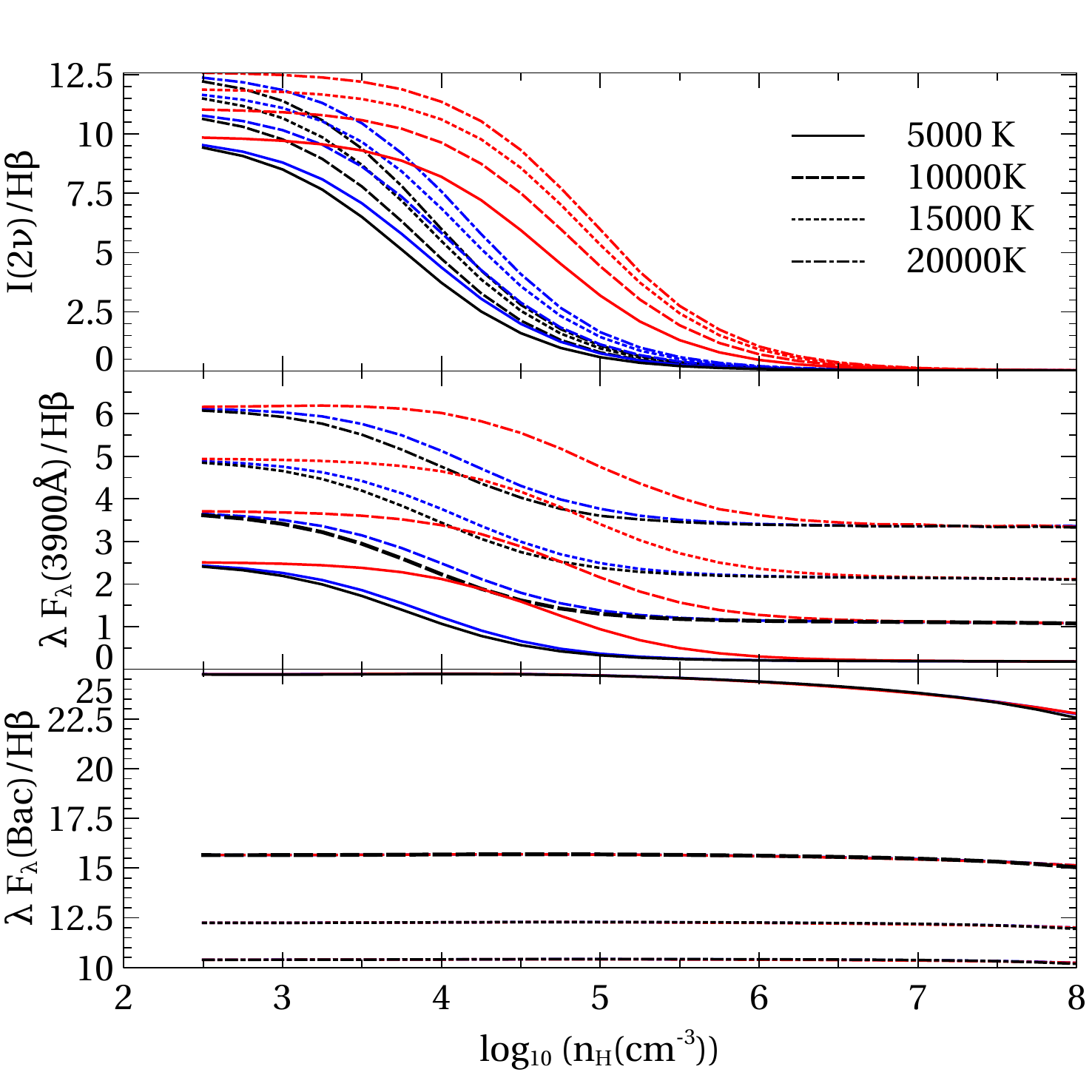}
\caption{\label{f:3pannels} Simulation of the ratio of selected fluxes to Case B H$\beta$.
Top: the total integrated two photon continuum intensity. Middle: The flux at 3900\AA.
Bottom: Balmer Jump. The different colors of the lines represent the different choice of
$l$-changing rates: PS-M/QM: black; SC: blue; SSC: red. 
}
\end{center}
\end{figure}

\section{Discussion}
\label{sec:disc}

The discussion above suggests that it may be possible to observationally test the $l$-changing
theory. Figure \ref{f:3pannels} shows three emission quantities, all relative to the Case B
H$\beta$. The top panel shows the integrated two photon continuum emission, the middle shows
the total flux at 3900\AA, while the bottom panel shows the flux at the head
of the BJ. These are plotted as a function of density and over the range of temperatures found
in nebulae. Predictions from three proposed $l$-changing collision theories are shown. 
Fig. \ref{f:3pannels} shows that the biggest differences in the two-photon flux resulting from
the various theories are found at densities $n_\text{H}\sim10^4\text{cm}^{-3}$, where the SSC
predictions are a factor $\sim3$ greater than the PS-M/QM ones. SC results are $\sim10$\% over
PS-M/QM. These densities have been observed in planetary nebulae. 

Many nebulae present complex geometries with observations often indicating that density
fluctuations are present \citep{Liu2012,Peimbert2013,Zhang2002b}. Using different
collisionally excited lines (CELs) to measure the density
lead typically to a factor $\sim$2 uncertainty in the value of the density
\citep{Peimbert2013,Fang2013,Fang2011}, although uncertainties can reach much higher values. 

An accurate determination of the temperature may be harder to obtain due to the well known
discrepancy between CEL and ORL abundances in planetary nebulae
\citep{GarciaRojas2007,Peimbert1967}, which can be expressed as
temperature variations. \citet{2016RMxAA..52..261F} discuss both this and the suggestion that
``kappa distribution'' electrons may be present. Other explanations have been proposed, such
as density fluctuations in the nebula \citep{Liu2000,McNabb2013} or by the effects of a binary
central star \citep{Jones2016}. Our two-photon spectrum should be associated with the ORL
temperatures, which can vary by a factor $\sim2$ depending on the observed lines. However, it
is not unusual to observe much larger discrepancies.

The observational challenge is great, but accurate $l$-changing collisional rates are needed
to obtain accurate lines emissivities.  As mentioned above, this cannot be done in the
laboratory.  Such a measurement would be specially desirable and would contribute to our
understanding of the basic atomic processes underlying nebular emission.

It should be noted that  accurate measurements of both temperature and density are needed to
make a decisive test of the various theories. 

\begin{figure}
\begin{center}
\includegraphics[width=0.4\textwidth]{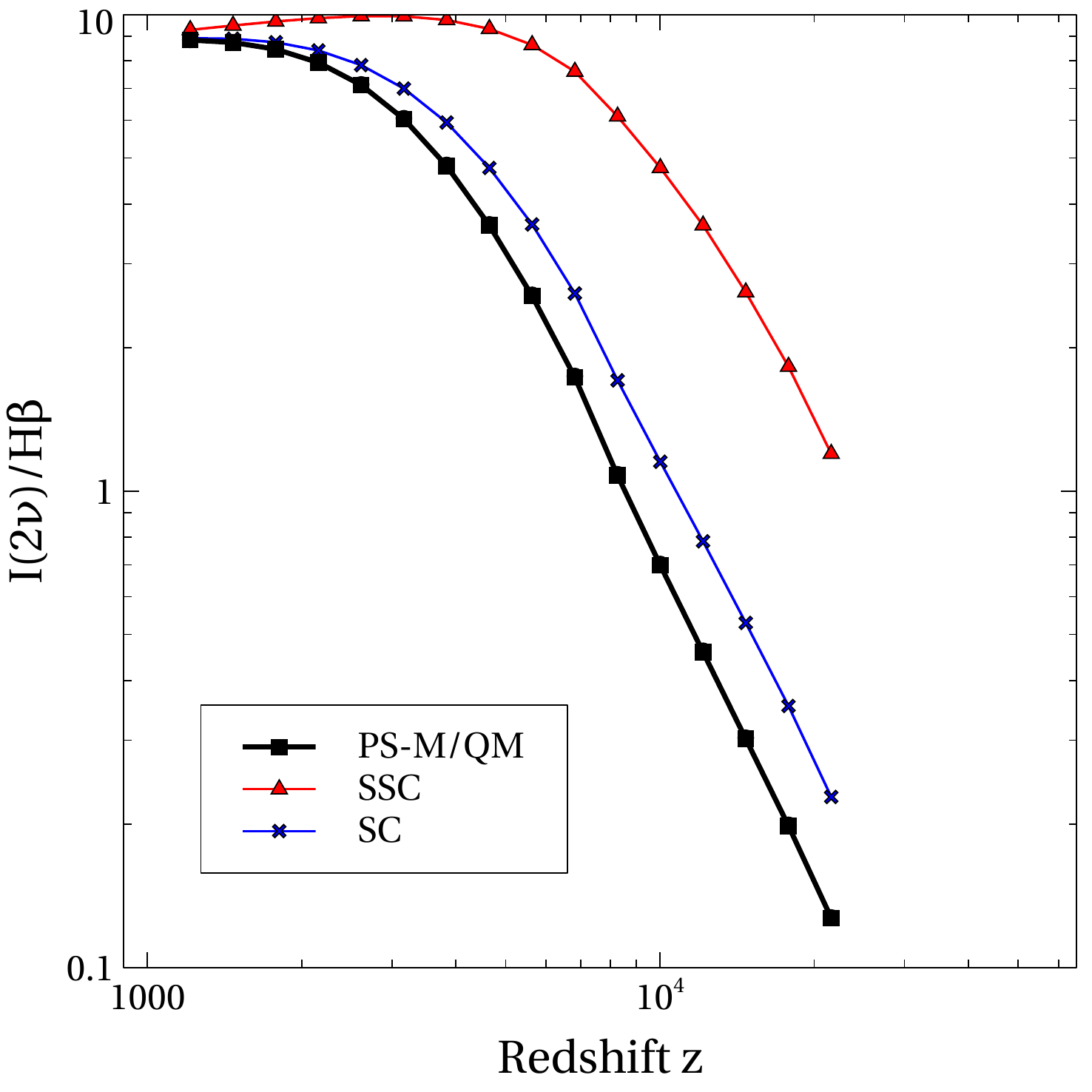}
\caption{\label{f:2nuredshift} The total integrated two-photon to H$\beta$ intensity ratio as
a function of z. As in Fig. \ref{f:15K}, the different lines represent the different theories
for the $l$-changing rates. Critical density $\rho_c=8.62\times10^{-30}\text{g cm}^{-3}$ and
$\Omega_bh^2=0.02230$ \citep{Planck2016}.  
}
\end{center}
\end{figure}

Finally, it is important to recall here that SSC rates have been used recently in CRR
calculations \citep{Chluba2016,Glover2014}, where a high precision of $\sim0.1$\% is required
(see Section \ref{sec:mot}). In Fig. \ref{f:2nuredshift}, we show the two-photon total
emission from steady state simulations due to the recombination of ionized hydrogen at the
temperature and density conditions of redshifts previous to the recombination epoch. Hydrogen
atoms are ionized using a ``laser'' source of 1.1Ry, as described in section \ref{sec:PN}.
These results suggest that significant errors in the two-photon emission can emerge by using
the SSC approximation. However, $l$-changing rate coefficients may not be relevant at $z<1000$
due to the predominance of radiative processes.

\section{Summary and conclusions}
\label{sec:sum}

The discrepancies between the various $l$-changing collisional theories have an impact on
accurate spectroscopic diagnostics of astrophysical plasmas. This paper examines their impact
on the two-photon continuum in nebulae. We provide updated $2s-2p$ $l$-changing rate
coefficients and \hi{} $n=2$ effective recombination coefficients. These last are only
slightly affected by the various $l$-changing collision theories at moderate to high
temperatures. However, these differences of 1\% could compromise tests of cosmological models
\citep{Chluba2006}. Effective recombination coefficients resulting from the three theories
differ at low temperatures ($T\sim 100$K) where high $n$-shell collisions compete with
radiative decay at densities where the system is not yet in local thermodynamic equilibrium.
We have also explored the effect that the various $l$-changing theories have on the two-photon
continuum spectrum of hydrogen in gaseous nebulae. 

These theoretical predictions of the two photon continuum could be compared with accurate
observations of simple objects, with reliable temperature and density determinations, to
determine which theory is correct. Astronomical observations are, at present, the only way to
conduct this test and settle the question. This is not a simple task, as
homogeneous object lacking small scale structure is needed. The planetary nebula A39 described in
\citet{Jacoby2001} is an example of it , however it is too low in density. Finding suitable
objects should be a high priority for future studies. We think these observations would be
challenging, but would offer a unique test for new atomic physics.

\section{Acknowledgments}

The authors acknowledge Prof. J. Chluba for for pointing out the importance of
atomic data on CRR simulations. We acknowledge support by NSF (1108928, 1109061, and 1412155), 
NASA (10-ATP10-0053, 10-ADAP10-0073, NNX12AH73G, and ATP13-0153), 
and STScI (HST-AR- 13245, GO-12560, HST-GO-12309, GO-13310.002-A, 
HST-AR-13914, and HST-AR-14286.001). MC has been supported by STScI (HST-AR-14286.001-A). PvH
was funded by the Belgian Science Policy Office under contract no.
BR/154/PI/MOLPLAN.

\bibliographystyle{mn2e}
\bibliography{LocalBibliography,LocalBibliography2.bib}

\begin{thebibliography}{}

\bibitem[\protect\citeauthoryear{{Baker} \& {Menzel}}{{Baker} \&
  {Menzel}}{1938}]{Baker1938}
{Baker} J.~G.,  {Menzel} D.~H.,  1938, \apj, 88, 52

\bibitem[\protect\citeauthoryear{{Brown} \& {Mathews}}{{Brown} \&
  {Mathews}}{1970}]{Brown1970}
{Brown} R.~L.,  {Mathews} W.~G.,  1970, \apj, 160, 939

\bibitem[\protect\citeauthoryear{{Chluba} \& {Ali-Ha{\"i}moud}}{{Chluba} \&
  {Ali-Ha{\"i}moud}}{2016}]{Chluba2016}
{Chluba} J.,  {Ali-Ha{\"i}moud} Y.,  2016, \mnras, 456, 3494

\bibitem[\protect\citeauthoryear{{Chluba} \& {Sunyaev}}{{Chluba} \&
  {Sunyaev}}{2006}]{Chluba2006}
{Chluba} J.,  {Sunyaev} R.~A.,  2006, \aap, 446, 39

\bibitem[\protect\citeauthoryear{{Chluba}, {Vasil} \& {Dursi}}{{Chluba}
  et~al.}{2010}]{Chluba2010}
{Chluba} J.,  {Vasil} G.~M.,    {Dursi} L.~J.,  2010, \mnras, 407, 599

\bibitem[\protect\citeauthoryear{{Fang} \& {Liu}}{{Fang} \&
  {Liu}}{2011}]{Fang2011}
{Fang} X.,  {Liu} X.-W.,  2011, \mnras, 415, 181

\bibitem[\protect\citeauthoryear{{Fang} \& {Liu}}{{Fang} \&
  {Liu}}{2013}]{Fang2013}
{Fang} X.,  {Liu} X.-W.,  2013, \mnras, 429, 2791

\bibitem[\protect\citeauthoryear{{Ferland}, {Binette}, {Contini}, {Harrington},
  {Kallman}, {Netzer}, {P{\'e}quignot}, {Raymond}, {Rubin}, {Shields},
  {Sutherland} \& {Viegas}}{{Ferland}
  et~al.}{2016}]{Ferland_Lexingtonbenchmark}
{Ferland} G.,  {Binette} L.,  {Contini} M.,  {Harrington} J.,  {Kallman} T.,
  {Netzer} H.,  {P{\'e}quignot} D.,  {Raymond} J.,  {Rubin} R.,  {Shields} G.,
  {Sutherland} R.,    {Viegas} S.,  2016, ArXiv e-prints, 1603.08902

\bibitem[\protect\citeauthoryear{{Ferland}, {Henney}, {O'Dell} \&
  {Peimbert}}{{Ferland} et~al.}{2016}]{2016RMxAA..52..261F}
{Ferland} G.~J.,  {Henney} W.~J.,  {O'Dell} C.~R.,    {Peimbert} M.,  2016,
  Revista Mexicana de Astronomia y Astrofisica, 52, 261

\bibitem[\protect\citeauthoryear{{Ferland}, {Porter}, {van Hoof}, {Williams},
  {Abel}, {Lykins}, {Shaw}, {Henney} \& {Stancil}}{{Ferland}
  et~al.}{2013}]{CloudyReview}
{Ferland} G.~J.,  {Porter} R.~L.,  {van Hoof} P.~A.~M.,  {Williams} R.~J.~R.,
  {Abel} N.~P.,  {Lykins} M.~L.,  {Shaw} G.,  {Henney} W.~J.,    {Stancil}
  P.~C.,  2013, Revista Mexicana de Astronomia y Astrofisica, 49, 137

\bibitem[\protect\citeauthoryear{{Garc{\'{\i}}a-Rojas} \&
  {Esteban}}{{Garc{\'{\i}}a-Rojas} \& {Esteban}}{2007}]{GarciaRojas2007}
{Garc{\'{\i}}a-Rojas} J.,  {Esteban} C.,  2007, \apj, 670, 457

\bibitem[\protect\citeauthoryear{{Glover}, {Chluba}, {Furlanetto}, {Pritchard}
  \& {Savin}}{{Glover} et~al.}{2014}]{Glover2014}
{Glover} S.~C.~O.,  {Chluba} J.,  {Furlanetto} S.~R.,  {Pritchard} J.~R.,
  {Savin} D.~W.,  2014, Advances in Atomic Molecular and Optical Physics, 63,
  135

\bibitem[\protect\citeauthoryear{{Guzm{\'a}n}, {Badnell}, {Williams}, {van
  Hoof}, {Chatzikos} \& {Ferland}}{{Guzm{\'a}n} et~al.}{2016}]{Guzman.I.2016}
{Guzm{\'a}n} F.,  {Badnell} N.~R.,  {Williams} R.~J.~R.,  {van Hoof} P.~A.~M.,
  {Chatzikos} M.,    {Ferland} G.~J.,  2016, \mnras, 459, 3498

\bibitem[\protect\citeauthoryear{{Guzm{\'a}n}, {Badnell}, {Williams}, {van
  Hoof}, {Chatzikos} \& {Ferland}}{{Guzm{\'a}n} et~al.}{2017}]{Guzman.II.2016}
{Guzm{\'a}n} F.,  {Badnell} N.~R.,  {Williams} R.~J.~R.,  {van Hoof} P.~A.~M.,
  {Chatzikos} M.,    {Ferland} G.~J.,  2017, \mnras, 464, 312

\bibitem[\protect\citeauthoryear{{Hummer} \& {Storey}}{{Hummer} \&
  {Storey}}{1987}]{Hummer1987}
{Hummer} D.~G.,  {Storey} P.~J.,  1987, \mnras, 224, 801

\bibitem[\protect\citeauthoryear{Irby, Rolfes, Makarov, MacAdam \& Syrkin}{Irby
  et~al.}{1995}]{Irby1995}
Irby V.~D.,  Rolfes R.~G.,  Makarov O.~P.,  MacAdam K.~B.,    Syrkin M.~I.,
  1995, Phys. Rev. A, 52, 3809

\bibitem[\protect\citeauthoryear{{Jacoby}, {Ferland} \& {Korista}}{{Jacoby}
  et~al.}{2001}]{Jacoby2001}
{Jacoby} G.~H.,  {Ferland} G.~J.,    {Korista} K.~T.,  2001, \apj, 560, 272

\bibitem[\protect\citeauthoryear{{Jones}, {Wesson}, {Garc{\'{\i}}a-Rojas},
  {Corradi} \& {Boffin}}{{Jones} et~al.}{2016}]{Jones2016}
{Jones} D.,  {Wesson} R.,  {Garc{\'{\i}}a-Rojas} J.,  {Corradi} R.~L.~M.,
  {Boffin} H.~M.~J.,  2016, \mnras, 455, 3263

\bibitem[\protect\citeauthoryear{{Liu}}{{Liu}}{2012}]{Liu2012}
{Liu} X.,  2012, in IAU Symposium Vol.~283 of IAU Symposium, {Atomic processes
  in planetary nebulae}.
pp 131--138

\bibitem[\protect\citeauthoryear{{Liu}, {Storey}, {Barlow}, {Danziger}, {Cohen}
  \& {Bryce}}{{Liu} et~al.}{2000}]{Liu2000}
{Liu} X.-W.,  {Storey} P.~J.,  {Barlow} M.~J.,  {Danziger} I.~J.,  {Cohen} M.,
    {Bryce} M.,  2000, \mnras, 312, 585

\bibitem[\protect\citeauthoryear{MacAdam, Crosby \& Rolfes}{MacAdam
  et~al.}{1980}]{MacAdam1980}
MacAdam K.~B.,  Crosby D.~A.,    Rolfes R.,  1980, Phys. Rev. Lett., 44, 980

\bibitem[\protect\citeauthoryear{MacAdam, Rolfes \& Crosby}{MacAdam
  et~al.}{1981}]{MacAdam1981}
MacAdam K.~B.,  Rolfes R.,    Crosby D.~A.,  1981, Phys. Rev. A, 24, 1286

\bibitem[\protect\citeauthoryear{MacAdam, Rolfes, Sun, Singh, Fuqua~III \&
  Smith}{MacAdam et~al.}{1987}]{MacAdam1987}
MacAdam K.~B.,  Rolfes R.~G.,  Sun X.,  Singh J.,  Fuqua~III W.~L.,    Smith
  D.~B.,  1987, Phys. Rev. A, 36, 4254

\bibitem[\protect\citeauthoryear{{McNabb}, {Fang}, {Liu}, {Bastin} \&
  {Storey}}{{McNabb} et~al.}{2013}]{McNabb2013}
{McNabb} I.~A.,  {Fang} X.,  {Liu} X.-W.,  {Bastin} R.~J.,    {Storey} P.~J.,
  2013, \mnras, 428, 3443

\bibitem[\protect\citeauthoryear{{Nussbaumer} \& {Schmutz}}{{Nussbaumer} \&
  {Schmutz}}{1984}]{NussbaumerSchmutz1984}
{Nussbaumer} H.,  {Schmutz} W.,  1984, \aap, 138, 495

\bibitem[\protect\citeauthoryear{{Osterbrock} \& {Ferland}}{{Osterbrock} \&
  {Ferland}}{2006}]{Osterbrock2006}
{Osterbrock} D.~E.,  {Ferland} G.~J.,  2006, {Astrophysics of gaseous nebulae
  and active galactic nuclei, 2nd.~ed.}.
Sausalito, CA: University Science Books

\bibitem[\protect\citeauthoryear{{Peimbert} \& {Peimbert}}{{Peimbert} \&
  {Peimbert}}{2013}]{Peimbert2013}
{Peimbert} A.,  {Peimbert} M.,  2013, \apj, 778, 89

\bibitem[\protect\citeauthoryear{{Peimbert}}{{Peimbert}}{1967}]{Peimbert1967}
{Peimbert} M.,  1967, \apj, 150, 825

\bibitem[\protect\citeauthoryear{{Pengelly}}{{Pengelly}}{1964}]{Pengelly1964}
{Pengelly} R.~M.,  1964, \mnras, 127, 145

\bibitem[\protect\citeauthoryear{{Pengelly} \& {Seaton}}{{Pengelly} \&
  {Seaton}}{1964}]{PengellySeaton1964}
{Pengelly} R.~M.,  {Seaton} M.~J.,  1964, \mnras, 127, 165

\bibitem[\protect\citeauthoryear{{P{\'e}quignot}}{{P{\'e}quignot}}{1986}]{Pequignot.D86Comparison-of-Photoionization-and-steady-shock}
{P{\'e}quignot} D.,  1986, in {Pequignot} D.,  ed., Model Nebulae {Comparison
  of Photoionization and steady shock models}.
p.~363

\bibitem[\protect\citeauthoryear{{P{\'e}quignot}, {Ferland}, {Netzer},
  {Kallman}, {Ballantyne}, {Dumont}, {Ercolano}, {Harrington}, {Kraemer},
  {Morisset}, {Nayakshin}, {Rubin} \& {Sutherland}}{{P{\'e}quignot}
  et~al.}{2001}]{Pequignot2001}
{P{\'e}quignot} D.,  {Ferland} G.,  {Netzer} H.,  {Kallman} T.,  {Ballantyne}
  D.~R.,  {Dumont} A.-M.,  {Ercolano} B.,  {Harrington} P.,  {Kraemer} S.,
  {Morisset} C.,  {Nayakshin} S.,  {Rubin} R.~H.,    {Sutherland} R.,  2001, in
  {Ferland} G.,  {Savin} D.~W.,  eds, Spectroscopic Challenges of Photoionized
  Plasmas Vol.~247 of Astronomical Society of the Pacific Conference Series,
  {Photoionization Model Nebulae}.
p.~533

\bibitem[\protect\citeauthoryear{{Planck Collaboration}, {Ade}, {Aghanim},
  {Arnaud}, {Ashdown}, {Aumont}, {Baccigalupi}, {Banday}, {Barreiro},
  {Bartlett} \& et al.}{{Planck Collaboration} et~al.}{2016}]{Planck2016}
{Planck Collaboration} {Ade} P.~A.~R.,  {Aghanim} N.,  {Arnaud} M.,  {Ashdown}
  M.,  {Aumont} J.,  {Baccigalupi} C.,  {Banday} A.~J.,  {Barreiro} R.~B.,
  {Bartlett} J.~G.,    et al. 2016, \aap, 594, A13

\bibitem[\protect\citeauthoryear{Porter, Ferland, MacAdam \& Storey}{Porter
  et~al.}{2009}]{2009MNRAS.393L..36P}
Porter R.~L.,  Ferland G.~J.,  MacAdam K.~B.,    Storey P.~J.,  2009, Monthly
  Notices of the Royal Astronomical Society: Letters, 393, L36

\bibitem[\protect\citeauthoryear{{Schirmer}}{{Schirmer}}{2016}]{2016PASP..128k4001S}
{Schirmer} M.,  2016, \pasp, 128, 114001

\bibitem[\protect\citeauthoryear{{Spitzer} Jr. \& {Greenstein}}{{Spitzer} \&
  {Greenstein}}{1951}]{Spitzer1951}
{Spitzer} Jr. L.,  {Greenstein} J.~L.,  1951, \apj, 114, 407

\bibitem[\protect\citeauthoryear{{Storey} \& {Sochi}}{{Storey} \&
  {Sochi}}{2015}]{2015MNRAS.446.1864S}
{Storey} P.~J.,  {Sochi} T.,  2015, \mnras, 446, 1864

\bibitem[\protect\citeauthoryear{Sun \& MacAdam}{Sun \&
  MacAdam}{1993}]{Sun1993}
Sun X.,  MacAdam K.~B.,  1993, Phys. Rev. A, 47, 3913

\bibitem[\protect\citeauthoryear{{Vrinceanu} \& {Flannery}}{{Vrinceanu} \&
  {Flannery}}{2001}]{Vrinceanu2001}
{Vrinceanu} D.,  {Flannery} M.~R.,  2001, \pra, 63, 032701

\bibitem[\protect\citeauthoryear{{Vrinceanu}, {Onofrio} \&
  {Sadeghpour}}{{Vrinceanu} et~al.}{2012}]{VOS2012}
{Vrinceanu} D.,  {Onofrio} R.,    {Sadeghpour} H.~R.,  2012, \apj, 747, 56

\bibitem[\protect\citeauthoryear{{Zhang} \& {Liu}}{{Zhang} \&
  {Liu}}{2002}]{Zhang2002b}
{Zhang} Y.,  {Liu} X.-W.,  2002, \mnras, 337, 499

\end{thebibliography}
\bsp

\label{lastpage}
\clearpage
\end{document}